\documentclass{edp-jp4}
\usepackage{graphicx}
\begin{document}
\newcommand {\bb}{\bibitem}
\newcommand {\be}{\begin{equation}}
\newcommand {\ee}{\end{equation}}
\newcommand {\bea}{\begin{eqnarray}}
\newcommand {\eea}{\end{eqnarray}}
\newcommand {\nn}{\nonumber}

\title{Scaling Relations in the Vortex State of Nodal Superconductors}

\author{Kazumi Maki}
\address{Max-Planck Institute for the Physics of Complex Systems,
N\"{o}thnitzer Str. 38, D-01187, Dresden, Germany}
\secondaddress{Department of Physics and Astronomy, University of Southern
California, Los Angeles, CA 90089-0484 USA}
\author{David Parker}
\sameaddress{1,2}
\author{Hyekyung Won}
\sameaddress{1}
\secondaddress{Department of Physics, Hallym University, Chuncheon 200-702, South Korea}

\date{\today}
\maketitle
\begin{abstract}

In contrast to multigap superconductors (e.g. MgB$_{2}$), the low-temperature
properties of nodal superconductors are dominated by nodal excitations.
Here we extend for a variety of nodal superocnductors the earlier work by
Simon and Lee and K\"ubert and Hirschfeld.  The scaling relations seen in
the thermodynamics and the thermal conductivity will provide an unequivocal
test of nodal superconductivity.

\end{abstract}
\section{Introduction}

Although nodal superconductors have been with us since 1979 \cite{1}, the
systematic study of the gap symmetry of these new superconductors began
only around 1994 with the establishment of 
d-wave symmetry of high-T$_{c}$ cuprate superconductors
through the angle resolved photoemission spectrum (ARPES) \cite{2}
and Josephson interferometry \cite{3,4}.  Unfortunately, however, these 
powerful techniques have not been applied to other nodal superconductors
like Sr$_{2}$RuO$_{4}$, heavy-fermion superconductors and organic 
superconductors.  

Since 2001 Izawa et al have succeeded in determining the gap functions
$|\Delta({\bf k})|$'s in Sr$_{2}$RuO$_{4}$ \cite{5}, CeCoIn$_{5}$ \cite{6}, 
$\kappa$-(ET)$_{2}$Cu(NCS)$_{2}$ \cite{7}, YNi$_{2}$B$_{2}$C \cite{8},
PrOs$_{4}$Sb$_{12}$, \cite{9,10}, and UPd$_{2}$Al$_{3}$ \cite{11,12}
through measurements
of the angle-dependent thermal conductivity in the vortex state.  These
experiments are only possible now since a) high-quality single crystals
of these compounds are now available, b) low-temperature facilities which
allow one to reach 1 - 0.1 K are available, and c)the necessary theoretical
development following the seminal paper by Volovik \cite{13}.

Indeed Volovik's approach has been extended in a variety of directions, as
reviewed in \cite{14}.  Also the angle dependent magentothermal conductivity
and the scaling relations \cite{15} in the vortex state will provide a 
crucial test of nodal superconductivity.  For example, the multigap
superconductors do not exhibit the scaling relations we are going to discuss
in general.  Therefore, if any given superconductor exhibits a scaling
relation discussed here, it is very likely that the material is a nodal
superconductor.  For example, the specific heat data of  Sr$_{2}$RuO$_{4}$
by Deguchi et al \cite{15} obeys the scaling relation given in \cite{16}.
Therefore the simplest choice of gap function in  Sr$_{2}$RuO$_{4}$ is 
the chiral f-wave superconductor as pointed out in \cite{5}.

The scaling relations in the vortex state in d-wave superconductors were
first proposed by Simon and Lee \cite{17}.  Then within the semiclassical
approximation, \`a la Volovik \cite{13} K\"ubert and Hirschfeld (KH) 
\cite{18} have succeeded in deriving the scaling function for the
quasiparticle density of states.  KH then calculated the thermal conductivity
in the scaling region \cite{19}.  An error in \cite{19}
was pointed out and corrected in \cite{20}.  However, in \cite{20} only the
asymptotic behavior of the thermal conductivity 
($T \ll <|{\bf v} \cdot {\bf q}|>$, where ${\bf v} \cdot {\bf q}$ is the
Doppler shift) has been worked out.  

In the following we shall derive the scaling relations for a class
of quasi-2D superconductors, where $|\Delta({\bf k})| = \Delta |f|$ and 
f = $\cos(2\phi), \sin(2\phi)$ (d-wave superconductor), $f = e^{\pm i\phi}\cos
\chi$ (chiral f-wave superconductor as in Sr$_{2}$RuO$_{4}$), $f = \cos(2\chi)$
(g-wave superconductor as in UPd$_{2}$Al$_{3}$ \cite{12}.)
These superconductors have
the same quasiparticle density of states as in d-wave superconductors
\cite{21}
\bea
N(E)/N_{0} &=& G(x)
\eea
where 
\bea
G(x)&=& \frac{2x}{\pi}K(x)\,\,\,\mathrm{for\,\, x \leq 1} \\
    &=& \frac{2}{\pi}K(x^{-1})\,\,\,\mathrm{for\,\, x > 1}.
\eea
where $x=|E|/\Delta$ and K(k) is the complete elliptic integral of the
second kind.  In particular for $|E| < 0.3 \Delta $ we have $G(E/\Delta) =
\frac{|E|}{\Delta}$.

As discussed elsewhere \cite{14}, all the nodal superconductors so far
discovered have G(E) $\sim |E|/\Delta$.  Then one can establish a variety
of scaling relations in the superclean limit \cite{20}, that is, for
$(\Gamma \Delta)^{1/2} < T,E;$ ${\bf v} \cdot {\bf q} < \Delta$ where
T, E, $\Delta$ and $\Gamma$ are the temperature, the quasiparticle energy,
the maximal value of the energy gap and the quasiparticle scattering rate
respectively.  Therefore the scaling relations provide another test for
nodal superconductivity.

\section{Quasiparticle density of states}

Let us limit ourselves to a class of quasi-2D systems with f listed in the
preceding section. As already noted we have $G(E) \simeq |E|/\Delta$ for
$E \ll \Delta$.  In the presence of a magnetic field we find 
\bea
G(E,{\bf H}) &=& <|E-{\bf v} \cdot {\bf q}|>\Delta^{-1}
\eea
where ${\bf v} \cdot {\bf q}$ is the Doppler shift and $\langle
\ldots \rangle$ means the average
over the Fermi surface and the vortex lattice.  When ${\bf H} \parallel 
{\bf c}$ in the class of quasi-2D systems, the average can be performed
analytically and we find \cite{18}
\bea
G(E,{\bf H}) &=& \frac{4}{\pi}\frac{\epsilon}{\Delta}g(E/\epsilon)
\eea
where
\bea
g(s)&=& \frac{\pi}{4}s(1+\frac{1}{2s^{2}}), s > 1 \\
    &=& \frac{3}{4}\sqrt{1-s^{2}}+\frac{1}{4s}(1+2s^{2})\arcsin(s),
\ \ \ \  s \leq 1
\eea
where $\epsilon = \frac{1}{2}v\sqrt{eH}$ and v is the Fermi velocity
within the ab plane.  The scaling function 
$\frac{\Delta}{\epsilon}G(E,{\bf H})$ is shown in Fig. 1.  As is readily
\begin{figure}[h]
\includegraphics[width=8cm]{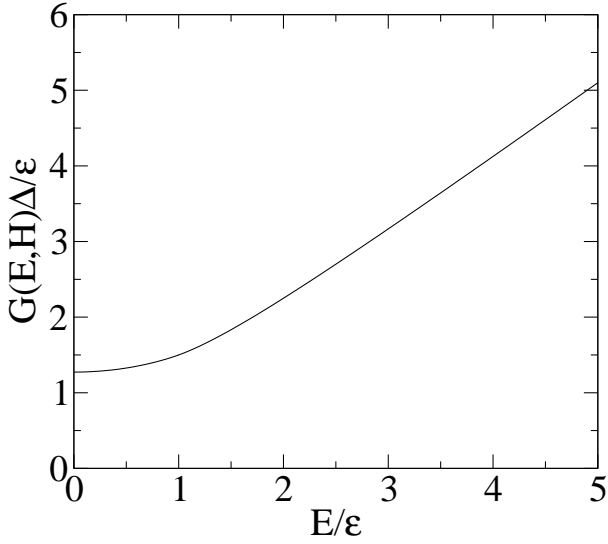}
\caption{The scaling function G(E, {\bf H})}
\end{figure}
seen $G(E,{\bf H})$ for ${\bf H} \parallel {\bf c}$ cannot discriminate
between different $|\Delta({\bf k})|$'s in the above class of nodal
superconductors.  Then the scaling function for the specific heat is given
by \cite{16}
\bea
C_{s}(T,{\bf H})/C_{s}(T,0) &=& F(T/\epsilon)
\eea
where 
\bea
F(T/\epsilon) &=& \frac{2}{9\pi \zeta(3)}(\frac{\epsilon}{T})^{2}
\int_{0}^{\infty}
ds \, s^{2}g(s) \mathrm{sech}^{2}\left(\frac{\epsilon s}{2T}\right) \\
& \simeq & 1 + \frac{\ln 2}{9 \zeta(3)}(\frac{\epsilon}{T})^{2},
\mathrm{for \,\, \epsilon/T \leq 1} \\
& \simeq & \frac{4\epsilon}{9\pi \zeta(3) T}[1 + 
\frac{1}{18}(\frac{\pi T}{\epsilon})^{2}+ \frac{7}{1800}
(\frac{\pi T}{\epsilon})^{4} + \ldots], \mathrm{for \,\, \epsilon/T > 1}
\eea
The scaling function and the experimental data for Sr$_{2}$RuO$_{4}$
\cite{15} are shown in Fig. 2.  As is seen readily the scaling function gives
an excellent description of the experimental data.  As noted in \cite{15},
\begin{figure}[h]
\includegraphics[width=8cm]{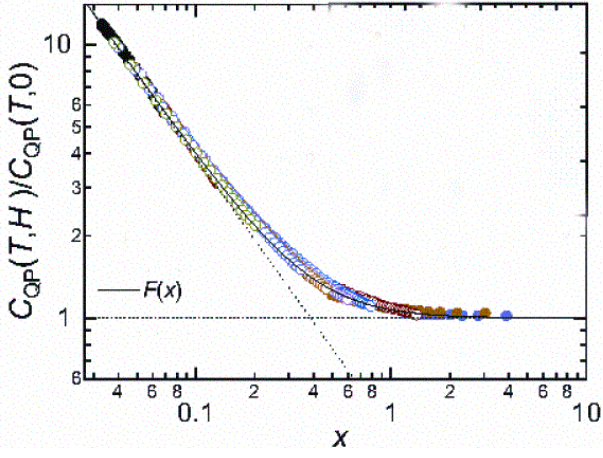}
\caption{The scaling function $F(T/\epsilon) = F(x)$ and Sr$_{2}$RuO$_{4}$
specific heat data from Ref. \cite{15}.}
\end{figure}
this clearly shows the superconductivity in Sr$_{2}$RuO$_{4}$ is consistent
with the chiral f-wave superconductor as discussed in \cite{23}.  On the
other hand, as noted in \cite{24}, this is incompatible with 
p-wave superconductivity.

Now when {\bf H} is rotated within the a-b plane with an angle $\phi$ from the 
a axis, we can discriminate between $f=\cos(2\phi)$ and $f=\sin(2\phi)$ (the
case of vertical nodes). We obtain for $f=\cos(2\phi)$
\bea
G(E,{\bf H}) & = &\frac{2}{\pi} \sum_{\pm}\left\langle 
\frac{\epsilon_{\pm}(\phi,\chi)}
{\Delta}G(\frac{E}{\epsilon_{\pm}(\phi,\chi)}\right\rangle \\
& = &  \frac{E}{\Delta}\left(1+ \frac{\epsilon^{2}}{2E^{2}}\right), 
\mathrm{for\,\,
\frac{\epsilon}{E} < 1}
\eea
\bea
= (\frac{2}{\pi})^{2}\frac{\epsilon}{\Delta}\sum_{\pm}(\sqrt{\frac{3}{2}
\pm \frac{1}{2} \sin(2\phi)} E\left(\frac{1}{\sqrt{\frac{3}{2} \pm 
\frac{1}{2} \sin(2\phi)}}\right) + \nonumber \\
\frac{1}{6}(\frac{E}{\epsilon})^{2}
\left(\frac{1}{\sqrt{\frac{3}{2} \pm \frac{1}{2} \sin(2\phi)}}\right)K
\left(\frac{1}{\sqrt{\frac{3}{2} \pm \frac{1}{2} 
\sin(2\phi)}}\right) + \ldots \\
\simeq \frac{4\epsilon}{\pi \Delta}(0.963+0.0205\cos(4\phi)+ \frac{1}{6}
(1.132 - 0.081 \cos(4\phi)) \times (\frac{E}{\epsilon})^{2} + \ldots) 
\mathrm{for \,\,\frac{\epsilon}{E} > 1}
\eea
where $\epsilon=\frac{1}{2}\tilde{v}\sqrt{eH}$ and 
$\tilde{v} = \sqrt{vv_{c}}$.  For f=$\sin(2\phi)$ we have the same formulas
as in Eqs. (12) and (13), except that $\cos(4\phi)$ in Eq.(14) should be
changed to $-\cos(4\phi)$.  Also the presence of the fourfold term in
the specific heat has been studied by Revaz et al \cite{25}.  They found no
fourfold term within an accuracy of 3\%.  This suggests strongly that the
thermal conductivity provides a more sensitive test of the gap symmetry.

For superconductors with horizontal nodes (e.g. f = $\sin \chi, \cos(2\chi),
\cos \chi$) the field configuration ${\bf H} \parallel b-c$ plane, with
$\theta$ the angle ${\bf H}$ makes from the c-axis, is more appropriate.  Then
we find \cite{12}
\bea
G(E,{\bf H}) &=& \frac{4}{\pi} \sum_{\pm}\left\langle \frac{\epsilon_{\pm}
(\theta,\phi,\chi)} {\Delta}G(\frac{E}{\epsilon_{\pm}(\theta,\phi,\chi)})\right\rangle \\
&=& \frac{E}{\Delta}(1+ \frac{\epsilon^{2}}{2E^{2}}), \mathrm{for\,\,
\frac{\epsilon}{E} < 1}
\eea
\bea
 = (\frac{2}{\pi})^{2}\frac{\epsilon}{\Delta}\sum_{\pm}\left(\sqrt{\frac{3}{2}
\pm \frac{1}{2} \sin(2\phi)} E(\frac{1}{\sqrt{\frac{3}{2} \pm 
\frac{1}{2} \sin(2\phi)}})\right) + \nonumber \\
\frac{1}{6}(\frac{E}{\epsilon})^{2}
\left(\frac{1}{\sqrt{\frac{3}{2} \pm \frac{1}{2} \sin(2\phi)}}\right)K
\left(\frac{1}{\sqrt{\frac{3}{2} \pm \frac{1}{2} \sin(2\phi)}}\right) \\
\eea
\bea
& \simeq & \frac{4\epsilon}{\pi \Delta}(0.963+0.0205\cos(4\phi)+ \frac{1}{6}
(1.132 - 0.081 \cos(4\phi))) \times (\frac{E}{\epsilon})^{2} + \ldots 
\mathrm{for \,\,\frac{\epsilon}{E} > 1}
\eea
\bea
\simeq \frac{4\epsilon}{\pi \Delta}\sqrt{x/2}(1 - 
\frac{1}{16}\sin^{2}\theta(\sin^{2}(\theta)+16\alpha^{2}\cos^{2}\theta
\sin^{2}\chi_{0})x^{-2} + \nn \\
\frac{1}{3}(\frac{E}{\epsilon})^{2}x^{-1}(1 +\frac{3}{16}\sin^{2}\theta(\sin^{2}(\theta)+16\alpha^{2}\cos^{2}\theta
\sin^{2}\chi_{0})x^{-2})) \mathrm{for \,\,\frac{\epsilon}{E} \geq 1}
\eea
where $\epsilon=v\sqrt{eH}$, $\alpha = v_{c}/v$ and 
$x = 1 +\cos^{2}\theta + 2\alpha^{2}\sin^{2}\theta \sin^{2} \chi_{0}$ and
$\chi_{0} = 0, \frac{\pi}{4}$ and $\frac{\pi}{2}$ for $f = \sin \chi, 
\cos(2\chi)$ and $\cos \chi$ respectively.  Therefore in the present 
configuration the angle dependent thermal conductivity can discriminate
different $\Delta({\bf k})$'s with horizontal nodes.

So far we have completely ignored the effect of impurity scattering.  As
already indicated the present analysis is valid in the superclean limit
\cite{14,20}, i.e. for $(\Gamma \Delta)^{1/2} < T,E$, ${\bf v} \cdot {\bf q} < 
\Delta$ where $\Gamma$ is the quasiparticle scattering rate in the normal state. Then the superclean limit appears to require $\Gamma/\Delta \leq 0.01$.

\section{Thermal conductivity}

In the past few years the angle dependent magnetothermal conductivity (ADMTC)
has proven itself the most powerful technique to probe the nodal structure
of the gap function $\Delta({\bf k})$.  Also in many cases the nodal 
structure of $\Delta({\bf k})$ is sufficient to deduce $\Delta({\bf k})$
itself.  We are concerned that much of the confusion and the 
controversy in the literature
regarding the gap functions in Sr$_{2}$RuO$_{4}$, PrOs$_{4}$Sb$_{12}$ and
UPd$_{2}$Al$_{3}$ may be largely 
due to a misunderstanding of Volovik's approach.  
References \cite{14,20} contain a detailed description of this approach. 
Generalizing the standard expression of the thermal conductivity given in
\cite{26,27}, the thermal conductivity of the class of nodal superconductors
in the vortex state is given by \cite{14}
\bea
\kappa_{zz} &=& \frac{n}{4mT^{2}}\int_{0}^{\infty}d \omega \, \omega^{2}
\left\langle\frac{h(\omega, {\bf H})}{\tilde{\Gamma}(\omega, {\bf H})}
\right\rangle \mathrm{sech}^{2}(\omega/2T)
\eea
where
\bea
h &=& \frac{1}{2}\left(1 + \frac{|\tilde{\omega}-{\bf v}\cdot{\bf q}|^{2}-
\Delta^{2}f^{2}}{|(\tilde{\omega}-{\bf v}\cdot{\bf q})^{2}-\Delta^{2}f^{2}|}
\right)
\eea
and
\bea
\tilde{\Gamma}&=& \mathrm{Im} \sqrt{(\tilde{\omega}-{\bf v}\cdot{\bf q})^{2}-
\Delta^{2}f^{2}}
\eea
Here $< ... >$ denotes the averages over the Fermi surface and vortex lattice
\cite{20}.  In the superclean limit $\tilde{\omega}$ is given by 
\bea
\tilde{\omega} &=& \omega + i\Gamma \left\langle\frac{|\tilde{\omega} -
{\bf v}\cdot {\bf q}|}{\sqrt{(\tilde{\omega}-{\bf v} \cdot {\bf q})^{2} -
\Delta^{2}f^{2}}}\right\rangle \\
& \simeq & \omega+i\Gamma G(\omega,{\bf H})
\eea
in the Born limit. And in the unitary limit we find 
\bea
\tilde{\omega} = \omega+i\Gamma G^{-1}(\omega,{\bf H})
\eea
where G($\omega,{\bf H}$) has been defined in Eq.(5).

First we limit ourselves to the quasi-2D systems in a magnetic field
{\bf H} $\parallel$ {\bf c}.  The in the Born limit we obtain
\bea
\kappa & = & \frac{n}{8mT^{2}\Gamma} \int_{0}^{\infty} d\omega\,\, \omega^{2}
\mathrm{sech}^{2}(\frac{\omega}{2T}= \frac{\pi^{2}nT}{12m\Gamma} = \frac{1}{2}
\kappa_{n}
\eea
where $\kappa_{n} = \frac{\pi^{2}nT}{6m\Gamma}$ is the thermal conductivity
in the normal state.  In particular Eq.(28) gives the scaling function
\bea
F_{B}(T/\epsilon) &=& \frac{\kappa(T,{\bf H})}{\kappa(T,0)} = 1
\eea
The last result agrees with the corresponding result given in Ref.\cite{19}
despite the use of a rather unphysical spatial average in this work.  In the 
unitary limit, on the other hand, we obtain
\bea
\kappa &=& \frac{n}{8m(T\Delta)^{2}\Gamma}\int_{0}^{\infty}d \omega \, \omega^{2}
<|\omega-{\bf v}\cdot{\bf q}|>^{2}\mathrm{sech}^{2}(\omega/2T) \\
&=& \frac{n}{8mT^{2}\Gamma}(\frac{\pi\epsilon}{4\Delta})^{2}\int_{0}^{\infty}d \omega \, \omega^{2} G^{2}(\omega/\epsilon)\mathrm{sech}^{2}(\omega/2T)
\eea
where $G(\omega/\epsilon)$ has already been defined in Eqs. 5 and 6.  This
has asymptotics
\bea
\kappa &=& \frac{7n \pi^4 T^{3}}{60m\Gamma\Delta^{2}}\left(
1+ \frac{5}{7}(\frac{\epsilon}{\pi T})^2 + \frac{15}{28}
(\frac{\epsilon}{\pi T})^{4} + \ldots \right), \,\,
\mathrm{for\,\, \epsilon \ll T} \\
&=& \frac{\pi^2 nT}{12m\Gamma}(\frac{\pi\epsilon}{4\Delta})^2\left(1+ \frac{7\pi^2}{15}(\frac{T}{\epsilon})^2 + \ldots\right), \,\, \mathrm{for\, \epsilon \gg T}
\eea
where $\epsilon=\frac{v\sqrt{eH}}{2}$.  Then the scaling function is given by
\bea
F_{u}(T/\epsilon) &=&  \frac{\kappa(T,{\bf H})}{\kappa(T,0)} = 
\frac{3}{2\pi^{2} T^{3}}\int_{0}^{\infty}d\omega \omega^{2} G^{2}(\omega/\epsilon) \mathrm{sech}^{2}(\omega/2T) \\
&=& \left(1+ \frac{5}{7}(\frac{\epsilon}{\pi T})^2 + \frac{15}{28}
(\frac{\epsilon}{\pi T})^{4} + \ldots \right), \,\,
\mathrm{for\,\, \epsilon \ll T} \\
&=& \frac{5}{112}(\frac{\epsilon}{\Delta})^2\left(1+ \frac{7\pi^2}{15}(\frac{T}{\epsilon})^2 + \ldots\right), \,\, \mathrm{for\, \epsilon \gg T}
\eea
These scaling functions are shown in Fig. 3.  In this figure we also include
\begin{figure}[h]
\includegraphics[width=7cm,angle=270]{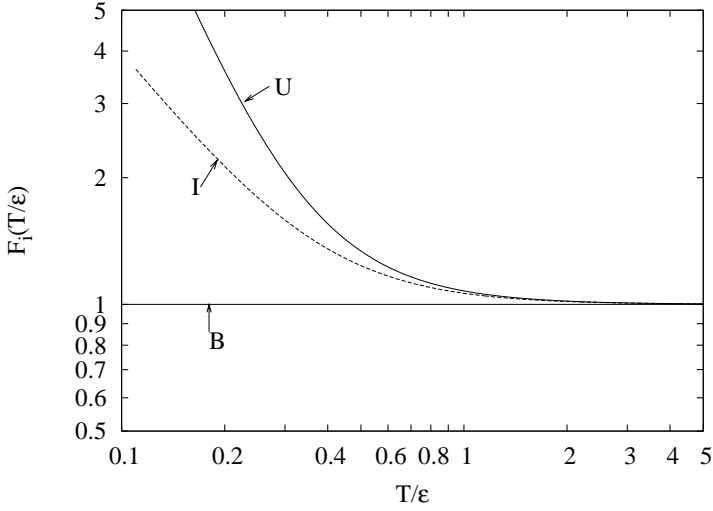}
\caption{The scaling functions F$_{I}$, F$_B$, and F$_U$}
\end{figure}
the scaling function when the inversion symmetry is broken in the impurity
scattering \cite{28}. $F_{I}(T/\epsilon)$ describes the thermal conductivity
data of the non-centrosymmetric triplet superconductor CePt$_{3}$Si by
Izawa et al \cite{29}.  The scaling function $F_{u}(T/\epsilon)$ is very
different from the one given in Ref. \cite{19} but describes consistently
the scaling behaviors of the thermal conductivity of UPt$_{3}$ as reported
by Suderow et al \cite{30}.

\section{Angle dependent thermal conductivity tensor}

Let us consider d$_{x^{2}-y^{2}}$-wave superconducivity as in the high-T$_{c}$
cuprates, CeCoIn$_{5}$ \cite{6} and $\kappa$-(ET)$_{2}$(NCS)$_{2}$ \cite{7}
in a magnetic field within the a-b plane.  Then the thermal conductivity tensors within the ab-plane are given by 
\bea
\kappa_{xx} &=& \kappa_{yy} = \frac{n}{8m\Gamma T^{2}} \int_{0}^{\infty}
d\omega \,\, \omega^{2} \left\langle \frac{4\epsilon(\phi,\chi)}{\pi \Delta}
G(\omega/\epsilon(\phi,\chi))\right\rangle ^{2} \mathrm{sech}^{2}(\omega/2T)
\eea
where $\epsilon(\phi,\chi) = \frac{\tilde{v}\sqrt{eH}}{2}(1 \pm \frac{1}{2}
\sin(2\phi) - \frac{1}{2}\cos(2\chi))^{1/2}$ and $\langle \ldots \rangle$
means the average over $\pm$ and over $\chi$.  Here we assumed the unitary
impurity scattering and the superclean limit in the present derivation. 
This gives the following asymptotics:
\bea
\kappa_{xx} &=& \kappa_{yy} = \frac{7nT}{60m\Gamma}(\frac{\pi T}{\Delta})^{2}\left(
1+ \frac{5}{7}(\frac{\epsilon}{\pi T})^2 + \frac{15}{28}
(\frac{\epsilon}{\pi T})^{4} + \ldots \right), \,\,
\mathrm{for\,\, \epsilon \ll T} \\
&=& \frac{4nT}{3m\Gamma}(\frac{\epsilon}{\Delta})^{2}[0.927 + 0.039 \cos(4\phi)
+ \frac{7}{15}(\frac{\pi T}{\epsilon})^{2}[1.090 - 0.055 \cos(4\phi)]
\mathrm{for\,\, \epsilon \gg T}
\eea
First of all, the present result is consistent with that in Ref. \cite{20}
for $T < \epsilon$.  On the other hand, for $T > \epsilon$ there is no fourfold
term.  In other words the present theory in the superclean limit cannot describethe fourfold symmetry in $\kappa_{xx}$ observed in YBCO for $ T > 14 K$
\cite{31,32,33}.  We have proposed the sign inversion of the fourfold term
for $T > \epsilon$ in the clean limit in \cite{34}.

Also the Hall conductivity in the present geometry is given by \cite{20}
\bea
\kappa_{xy} &=& \frac{n}{8m(T\Delta)^{2}\Gamma}\int_{0}^{\infty}
d\omega \,\, \omega^{2} \left\langle \sin(2\phi^{`})|\omega - {\bf v} \cdot
{\bf q}| \right\rangle  \left\langle |\omega - {\bf v} \cdot
{\bf q}| \right\rangle \mathrm{sech}^{2}(\omega/2T)
\eea
Further, we find
\bea
\left\langle \sin(2\phi^{`})|\omega - {\bf v} \cdot
{\bf q}| \right\rangle &=& -\sin(2\phi) \frac{\epsilon^{2}}{2|\omega|}, \mathrm{for\,\,
\epsilon \ll |\omega|} \\
& \simeq & -\sin(2\phi)\frac{4\epsilon}{\pi\Delta}(0.535 - 
(\frac{\omega}{\epsilon})^{2}0.14192)\, \mathrm{for\,\, \epsilon \gg |\omega|}
\eea
Inserting these into Eq.(40) we find 
\bea
\kappa_{xy} &=& -\sin(2\phi) \frac{\pi^{2}nT}{24m\Gamma}(\frac{\epsilon}{\Delta})^{2}
(1 + \frac{3}{2}(\frac{\epsilon}{\pi T})^{2}, 
\mathrm{for \,\, \epsilon \ll T} \\
&=& -\frac{4nT}{3m\Gamma}(\frac{\epsilon}{\Delta})^{2}\sin(2\phi)(0.5152 +
0.011 \cos(4\phi))-\frac{7}{30}(\frac{\pi T}{\epsilon})^{2}(0.213 + .0607
\cos(4\phi))\,\,\,\,\,\,\,\,\,\,\,\,\,\,\,\,\,\,\,\,\,\,\,\,\,\,\,\,\,\,\
\,\,\,\,\,\,\,\,\,\,\,\,\,\,\,\,\,\,\,\, \mathrm{for \epsilon \gg T}
\eea
Therefore in the superclean limit $\kappa_{xy} \sim -\sin(2\phi)H$ independent
of $\epsilon/T$.  Also as $\frac{T}{\epsilon}$ increases the coefficient
of -$\sin(2\phi) H$ decreases almost 40\%.  The present result appears to be
consistent with the Hall conductivity data of YBCO reported by Oca\~{n}a and
Esquinazi \cite{35} for $\epsilon/\Delta < 1$.  Also in the superclean limit 
the sign of the Hall conductivity is the same for all T as long as 
$T < \Delta(T)$.

\section{Concluding Remarks}

We have shown a) the thermal conductivity in nodal superconductors for 
$T < 0.3 T_{c}$ is dominated by the quasiparticles or nodal excitations,
b) the quasiparticles in the vortex state are accurately described in terms
of the semiclassical approximation.  Thus the angle dependent magneto-thermal
conductivity provides a powerful tool to determine the nodal structure
of the gap function as demonstrated by a series of experiments by Izawa et al
\cite{5,6,7,8,9,10,11,12}.  Also in most cases the nodal structure of the
gap function is adequate to deduce $|\Delta({\bf k})|$ itself.  In addition
we have shown that all these model superconductors exhibit a variety of
scaling relations.  We have proposed scaling relations for PrOs$_{4}$Sb$_{12}$
\cite{36}.  Furthermore, from the unusual scaling relation seen in the thermal
conductivity in CePt$_{3}$Si we can deduce anomalous impurity scattering in
this system lacking crystalline inversion symmetry.  Indeed the scaling
relations in nodal superconductors provide a unique way to characterize
this new class of superconductors.

\end{document}